\documentclass[pdflatex,referee,sn-mathphys-num]{sn-jnl}

\usepackage{graphicx}%
\usepackage{multirow}%
\usepackage{amsmath,amssymb,amsfonts}%
\usepackage{amsthm}%
\usepackage{mathrsfs}%
\usepackage[title]{appendix}%
\usepackage{xcolor}%
\usepackage{textcomp}%
\usepackage{manyfoot}%
\usepackage{booktabs}%
\usepackage{algorithm}%
\usepackage{algorithmicx}%
\usepackage{algpseudocode}%
\usepackage{listings}%
\usepackage{dcolumn}
\usepackage{multirow}
\usepackage{makecell}
\usepackage{bm}

\def\mrg#1{{\color{red}{#1}}}  

\usepackage[normalem]{ulem}
\begin{document}

\title[Article Title]{Atomic determination of the nuclear quadrupole moment $\mathrm{Q}(^{209}{\rm Bi})$ using the multi-configuration Dirac-Hartree-Fock method}

\author*[1]{\fnm{Jiguang} \sur{Li}}\email{li\_jiguang@iapcm.ac.cn}
\author[2]{\fnm{Jacek} \sur{Biero{\'n}}}
\author[3]{\fnm{Michel} \sur{Godefroid}}
\author[4]{\fnm{Per} \sur{Jönsson}}

\affil*[1]{\orgname{Institute of Applied Physics and Computational Mathematics},
\orgaddress{\city{Beijing}, \postcode{100088}, \country{China}}}
\affil[2]{\orgdiv{Instytut Fizyki Teoretycznej}, \orgname{Uniwersytet Jagiello{\'n}ski}, \city{Krak{\'o}w}, \country{Poland}}
\affil[3]{\orgdiv{Spectroscopy, Quantum Chemistry and Atmospheric Remote Sensing}, \orgname{Universit{\'e} libre de Bruxelles}, \orgaddress{\city{Brussels}, \postcode{1050}, \country{Belgium}}}
\affil[4]{\orgdiv{Department of Materials Science and Applied Mathematics}, \orgname{Malm\"o University}, \orgaddress{\city{Malm\"o}, \postcode{S-20506}, \country{Sweden}}}

\abstract{
The multiconfiguration Dirac-Hartree-Fock method implemented in the Grasp2018 package was employed to calculate the magnetic dipole hyperfine interaction constants and electric field gradients of levels in the ground configuration of the neutral bismuth atom. Combining the calculated electric field gradient of the ground state with the measured electric quadrupole hyperfine interaction constant, we extracted the nuclear quadrupole moment for the $^{209}$Bi isotope, $\mathrm{Q}(^{209}\textbf{Bi}) = -422(22)$~mb. This value, together with other results obtained from atomic- and molecular-structure calculations, created the ``world average" nuclear quadrupole moment of this isotope, $\mathrm{Q}(^{209}\textbf{Bi}) = -420(17)$~mb.
}

\maketitle
%
\section{Introduction}
In 2021 Barzakh \textit{et al.} reported measured magnetic dipole ($A$) and electric quadrupole~($B$) hyperfine interaction constants of the ground state $6p^3~^4\!S_{3/2}$ of neutron-deficient bismuth isotopes $^{187,188,189,191}$Bi, using the in-source resonance-ionization spectroscopy techniques at ISOLDE~(CERN)~\cite{Barzakh2021}. The nuclear quadrupole moments ($\mathrm{Q}$) of these isotopes can be determined if the accurate quadrupole moment is available for the reference isotope $^{209}$Bi, through $\mathrm{Q}(^{A}\text{Bi})/\mathrm{Q}(^{209}\text{Bi})=B(^{A}\text{Bi})/B(^{209}\text{Bi})$. Unfortunately, however, the existing values of $\mathrm{Q}(^{209}\text{Bi})$ were inconsistent with each other at that time. In fact, the electric quadrupole moment Q($^{209}$Bi) has a complicated history, suffering from a longstanding atomic-molecular discrepancy, and characterized by a wide spread of results from nuclear theory, muonic, pionic, etc. measurements; see the two series of reviews by Stone~\cite{Stone2016} and Pyykk{\"o}~\cite{Pyykko2018}. The inconsistencies stimulated a large-scale collaboration to revisit Q($^{209}$Bi), which involved several groups from Australia, Belgium, China, Finland, France, Lithuania, Poland, Russia and Sweden, to carry out independent atomic- and molecular-structure calculations, based on different methods with various computational models. Combining measured values of the $B$ constants of several electronic states either in atomic systems such as the neutral $^{209}$Bi atom or the $^{209}$Bi$^+$ ion, or in several diatomic molecules (BiN, BiP, BiF, BiI, BiO, BiCl) with calculated electric field gradient~(EFG) values for those particular systems, a ``world  average" $\mathrm{Q}(^{209}\text{Bi}) = -420(17)$~mb was obtained with a standard deviation of a sample of 33 theoretical results of $\mathrm{Q}(^{209}\text{Bi})$~\cite{Barzakh2021}. Some of the results were published~\cite{Skripnikov2021, Wilman2021, Dognon2023}. In this paper, we reported the contributions to the ``world average" $\mathrm{Q}(^{209}\text{Bi})$ value from the multi-configuration Dirac-Hartree-Fock~(MCDHF) calculations, combined with relativistic configuration interaction (RCI) by using the Grasp2018 package~\cite{FroeseFischer2019a, Jonsson2023} based on the variational approach~\cite{Jonsson2022}.

\section{Thoretical Method}
\subsection{Hyperfine interaction }
The Hamiltonian of the hyperfine interaction reads~\cite{Schwartz1955a, Johnson2007}
\begin{equation}
\label{Hamiltonian_hfs}
H_{\rm hfs} = \sum_{k} \mathcal{\bm T}^{(k)} \cdot {\bf T}^{(k)} \,,
\end{equation}
where $\mathcal{\bm T}^{(k)}$ and ${\bf T}^{(k)}$ are spherical tensor operators of rank $k$ in the nuclear and electronic spaces, respectively. The $k=1$ represents the magnetic dipole hyperfine interaction and $k=2$ the electric quadrupole hyperfine interaction.  Higher-order terms beyond $k=2$ are neglected due to their fractional contribution to hyperfine splittings. According to the first-order perturbation theory, the corrections from hyperfine interaction to energies of atomic states are given by
\begin{equation}
\label{First_hfs}
W^{(1)}_{\rm hfs} = \langle \Upsilon \Gamma I J F M_{F} |\sum_k \mathcal{\bm T}^{(k)} \cdot {\bf T}^{(k)}| \Upsilon \Gamma I J F M_{F} \rangle \,,
\end{equation}
where the unperturbed wavefunction is coupled by the nuclear wavefunction $| \Upsilon I M_I \rangle$ and electronic wavefunction $|\Gamma J M_J \rangle$ through
\begin{equation}
|\Upsilon \Gamma I J F M_F \rangle = \sum_{M_I,M_J} \langle I J M_I M_J | I J F M_F \rangle |\Upsilon I M_I \rangle |\Gamma J M_J \rangle \, . 
\end{equation}
Here, $I$ is the nuclear angular momentum, $M_I$ stands for its component along the quantization axis, and $\Upsilon$ represents other quantum numbers designating the nuclear state. $J$, $M_J$ and $\Gamma$ are the corresponding quantum numbers for the electronic state. With assistance of the angular momentum theory~\cite{Brink1968}, matrix elements in Eq. (\ref{First_hfs}) are reduced to
\begin{align}
\label{reduced_matrix_element}
&\langle \Upsilon \Gamma I J F M_F |\mathcal{\bm T}^{(k)} \cdot {\bf T}^{(k)}| \Upsilon \Gamma I J F M_F \rangle \nonumber \\
=& (-1)^{F-I-J} \sqrt{(2I+1)(2J+1)} \begin{Bmatrix}
                                                             I   &  I    &   k    \\
                                                            J  &  J    &   F 
                                                   \end{Bmatrix}\langle \Upsilon I ||\mathcal{\bm T}^{(k)}|| \Upsilon I \rangle \langle \Gamma J ||{\bf T}^{(k)}|| \Gamma J \rangle
\end{align}

Introducing the coefficient $\mathfrak{M}$,
\begin{equation}
\mathfrak{M}(I,J,F,k) = \frac{(-1)^{F-I-J}\begin{Bmatrix}
                                                             I   &  I    &   k    \\
                                                            J   &  J    &   F 
                                                   \end{Bmatrix}}
                         {\begin{pmatrix}
                                 I   &  k    &   I    \\
                                 -I  &  0    &   I 
                           \end{pmatrix}\begin{pmatrix}
                                                             J   &  k   &   J    \\
                                                            -J   &  0   &   J 
                                                 \end{pmatrix}}  \, ,
\end{equation}
the first-order correction from hyperfine interactions (Eq. (\ref{First_hfs})) is simplified to 
\begin{equation}
W_{\rm hfs}^{(1)} = \sum_k \mathfrak{M}(I, J, F, k) \langle \Upsilon I I |\mathcal{T}^{(k)}_0| \Upsilon I I \rangle \langle \Gamma J J |{T}^{(k)}_0| \Gamma J J \rangle \,.
\end{equation}
The first two coefficients $\mathfrak{M}(I,J,F,k)$ in terms of $K=F(F+1)-I(I+1)-J(J+1)$ are the following:
\begin{align}
\mathfrak{M}(I,J,F,1) =& \frac{K}{2IJ} \,,\\
\mathfrak{M}(I,J,F,2) =& \frac{6K(K+1) - 8I(I+1)J(J+1)}{4I(2I-1)~J(2J-1)}  \,.
\end{align}
Furthermore, magnetic dipole and electric quadrupole hyperfine interaction constants are defined as~\cite{Schwartz1955a}
\begin{align}
A &= \frac{1}{IJ} \langle \Upsilon I I |\mathcal{T}^{(1)}_0| \Upsilon I I \rangle \langle \Gamma J J |{T}^{(1)}_0| \Gamma J J \rangle = \frac{\mu_I}{IJ} \langle \Gamma J J |{T}^{(1)}_0| \Gamma J J \rangle \,, \\
B &= 4 \langle \Upsilon I I |\mathcal{T}^{(2)}_0| \Upsilon I I \rangle \langle \Gamma J J |{T}^{(2)}_0| \Gamma J J \rangle = 2 \mathrm{Q} \langle \Gamma J J |{T}^{(2)}_0| \Gamma J J \rangle  \, .
\end{align}
Here, $\mu_I = \langle \Upsilon I I |\mathcal{T}^{(1)}_0| \Upsilon I I \rangle$ is the nuclear dipole moment and $\mathrm{Q}=\langle \Upsilon I I |\mathcal{T}^{(2)}_0| \Upsilon I I \rangle$ the nuclear quadrupole moment. The electronic matrix elements are reduced to
\begin{align}
\langle \Gamma JJ |T^{(k)}_0| \Gamma JJ \rangle =&\sqrt{2J+1}  \begin{pmatrix}
                                                                                   J  &  k   &   J    \\
                                                                                  -J  &  0  &   J  
                                                                     \end{pmatrix} \langle \Gamma J || {\bf T}^{(k)} || \Gamma J \rangle \nonumber \\
=&  \frac{\sqrt{2J+1}~(2J)!}{\sqrt{(2J-k)!(2J+k+1)!}} \langle \Gamma J || {\bf T}^{(k)} || \Gamma J \rangle \,.
\end{align}
Therefore, the $A$ and $B$ hyperfine interaction constants can be rewritten to
\begin{subequations}\label{HFS_cons}
\begin{gather}
A =  \frac{\mu_I}{I} \frac{1}{\sqrt{J(J+1)}} \langle \Gamma J ||{\bf T}^{(1)}|| \Gamma J \rangle \label{A_cons} \,, \\
B = 2 \mathrm{Q} \sqrt{\frac{J(2J-1)}{(J+1)(2J+3)}} \langle \Gamma J ||{\bf T}^{(2)}|| \Gamma J \rangle \label{B_cons} \,.
\end{gather}
\end{subequations}
Specifically, for an $N$-electron atomic system, in atomic units~(a.u.), 
\begin{subequations}
\begin{gather}
{\bf T}^{(1)} = \sum_{\nu=1}^{N} {\bf t}^{(1)}_{\nu} =  \sum_{\nu=1}^{N} -\frac{i\alpha~({\bm \alpha}_{\nu} \cdot {\mathbf l}_{\nu} {\bf C}^{(1)}_{\nu})}{r^{2}_{\nu}}      \,,\\
{\bf T}^{(2)} = \sum_{\nu=1}^{N} {\bf t}^{(2)}_{\nu} = \sum_{\nu=1}^{N} - \frac{{\bf C}^{(2)}_{\nu}}{r^3_{\nu}}  \,,
\end{gather}
\end{subequations}
where $\alpha$ is the fine-structure constant, ${\bm \alpha}$ is the Dirac matrix, ${\mathbf l}$ is the orbital angular momentum operator, and ${\bf C}^{(k)}$ is a spherical tensor. 

The nuclear quadrupole moment $\mathrm{Q}$~(in barn) can be extracted from an experimental determination of the electric quadrupole hyperfine interaction constant~(in MHz), if the electric field gradient~(EFG) $q$~(in a.u.) is provided by an atomic-structure calculation, that is,
\begin{equation}
\mathrm{Q} = 4.255957 \times 10^{-3} \left( \frac{B_{\text{exp}}}{q} \right) \,.
\end{equation}
Comparing with Eq.~(\ref{B_cons}), it is clear that 
\begin{equation}
q= \sqrt{\frac{4J(2J-1)}{(J+1)(2J+3)}} \, \langle \Gamma J ||{\bf T}^{(2)}|| \Gamma J \rangle \,.
\end{equation}
Similarly, the nuclear dipole moment can be derived from the measured magnetic dipole hyperfine interaction constant if the magnetic field within the nucleus produced by the electrons~(MFE) $A_{\text{el}}$ is known. In our work, $A_{\text{el}}$ is defined as 
\begin{equation}
A_{\text{el}} = \frac{1}{\sqrt{J(J+1)}} \langle \Gamma J ||{\bf T}^{(1)}|| \Gamma J \rangle \,.
\end{equation}

\subsection{MCDHF method}
Calculations of EFGs and MFEs are based on wave functions of electronic states $|\Gamma J M_J \rangle$ that are approximated by atomic state functions~(ASFs). In this work, we employed the MCDHF approach to obtain the ASFs. Details can be found in the monograph by Grant~\cite{Grant2007}, or in the review papers~\cite{Fischer2016a, Jonsson2022}. In the framework of the MCDHF method, an ASF is a linear combination of configuration state functions~(CSFs) with the same total angular momentum $J$, the same quantum number $M_J$ for its component along the quantization axis, and the same parity $P$
\begin{equation}\label{ASF}
|\Gamma J M_J \rangle= \sum_{\nu=1}^{N_{\text{CSF}}} c^{\Gamma J}_{\nu} |\gamma_{\nu} J M_J \rangle \,,
\end{equation}
where letter $P$ is omitted for convenience,
and $N_{\text{CSF}}$ represents the total number of CSFs. Each CSF is composed of Slater determinants, built from relativistic one-electron orbitals. These orbitals, together with the mixing coefficients $c^{\Gamma J}_{\nu}$ in Eq.~(\ref{ASF}) are obtained by solving the MCDHF  equations self-consistently to reach a stationary point of the expectation value of the Hamiltonian. In the Dirac-Coulomb approximation for an $N$-electron atomic system, the Hamiltonian is expressed as 
\begin{equation}
\mathcal{H}_{\text{DC}} = \sum_{i=1}^N \left[ c {\bm \alpha}_i \cdot {\bf p}_i + c^2(\beta_i -1) + V_{\text{nuc}}(r_i) \right] + \sum_{j>i}^{N} \frac{1}{r_{ij}} \,,
\end{equation}
where $c$ is the light speed, ${\bm \alpha}$ and $\beta$ are the Dirac matrices, and $V_{\text{nuc}}$ is the nuclear potential produced by the nucleus, described with a two-parameter Fermi nuclear charge distribution model~\cite{Parpia1992}.

Once an orbital set is obtained, relativistic configuration interaction~(RCI) computations may be performed to systematically account for electron correlations. 
At this stage, only the mixing coefficients are variational. The Breit interaction
\begin{equation}
\mathcal{H}_{\text{B}} = - \sum_{j>i=1}^{N} \frac{1}{2r_{ij}} \left[ {\bm \alpha}_i \cdot {\bm \alpha}_j + \frac{({\bm \alpha}_i \cdot \mathbf{r}_{ij})({\bm \alpha}_j \cdot \mathbf{r}_{ij})}{r_{ij}^2} \right] \,,
\end{equation}
can also be added to the Dirac-Coulomb Hamiltonian in these RCI calculations.

\section{Computational Models and Results}
The neutral bismuth atom has 83 electrons with an open $p$ subshell in its ground configuration [Xe]$4f^{14}5d^{10}6s^26p^3$. For such an atomic system it is challenging to capture electron correlation effects and estimate the computational uncertainty. 
In this work, we adopted a systematic active set approach~\cite{Fischer2016a,Jonsson2023,CAS,RAS}, which is commonly used to deal with open-shell systems~\cite{BrI,Bieetal:2018a,Papetal:2021a}.
Generally, the interaction between configurations belonging to the same complex is expected to be large~\cite{LayBah:62a,Fischer2016a}. 
For the present $n=6$ complex case, test calculations reveal a large interaction between $6s^26p^3$ and $6s^26p6d^2$ (in non-relativistic notation). Therefore, these two configurations define the multi-reference (MR) set~\cite{Li2012, Jonsson2022}.
The electron correlation effects
are captured by including, in the ASF expansion~(\ref{ASF}), the configuration state functions (CSFs) that are generated by single, double, triple, or multiple substitutions of selected orbitals of the MR configurations with correlation orbitals~\cite{Li2012,Filippin2017,Papetal:2021a}. To simplify the notation,  the atomic core [Xe]$4f^{14}5d^{10}$ is omitted in further text, and only the valence subshells are presented.

In the first stage of our calculations we generated the orbital set, which was composed of spectroscopic and correlation orbitals~\cite{Jonsson2023}. Spectroscopic orbitals with the same principal quantum number are spatially clustered and have similar one-electron energies, while correlation orbitals do not have such properties because their orbital energies have no physical meaning~\cite{Bieron2009a}. 
We started from optimization of the occupied orbitals in the MR configurations $\{ 6s^26p^3,6s^26p6d^2\}$ to minimize the configuration average energy of $6s^26p^3$ using the MCDHF method~\cite{Jonsson2022, Jonsson2023}. 
These orbitals, with the exception of $6d$, were treated as spectroscopic. To minimize the configuration average energy, the extended optimal level scheme~\cite{Grant2007, Dyall1989} was used in the self-consistent-field (SCF) procedures~\cite{FroeseFischer1997}. This step is labeled as MR-DHF.

The shapes of correlation orbitals, which largely depend on the correlation effects that they describe \cite{Godetal:98a,Veretal:2010a}, influence to a large extent the computational efficiency and accuracy of the theoretical results. Therefore, we tailored the CSF expansion used for the correlation orbital optimization by targeting  the valence-valence (VV) and core-valence (CV) correlation through the inclusion of the corresponding substitutions.
The VV CSFs account for the correlation between electrons in the $6s$ and $6p$ valence subshells, while the CV CSFs account for the correlation between an electron in a core subshell and another one in the valence. The VV CSFs were generated by single (S) and double (D) substitutions from the $6s$, $6p$ and $6d$ valence orbitals in the MR configurations to correlation orbitals, while the CV CSFs by single and restricted double (SrD) substitutions from the orbitals only in the $4s^24p^64d^{10}4f^{14}5s^26p^65d^{10}6s^26p^3$ configuration. The restriction in those double substitutions refers to that only one electron from the $4s4p4d4f5s6p5d$ core orbitals can be promoted at a time. 
The set of correlation orbitals was augmented layer by layer~\cite{Jonsson2022}, and each layer consists of orbitals with different spatial symmetries, that is:~$s,~p,~d,~\dots$~up to $i$. While correlation orbitals were successively added, the atomic core shells with $n=4$ and $n=5$ were successively opened for substitutions, starting from the second layer. This allowed us to monitor the convergence of atomic properties concerned. 
In each SCF procedure, only the correlation orbitals of the last added layer were variationally optimised, and all others were kept frozen. This is the essence of the layer by layer approach~\cite{Jonsson2022}. Table~\ref{HFS_convergence} shows which substitutions from occupied orbitals (referred to as the active orbitals) in a given reference configuration were performed (allowed) to a specified set of correlation orbitals. For instance, the 3rd line in the first two columns represents the substitutions from $5s5p5d6s6p$ orbitals in the $5s^25p^65d^{10}6s^26p^3$ configuration and $6s6p6d$ in $6s^26p6d^2$ to $2spdfg$, i.e.~to ten correlation orbitals -- two of each of $s, \, p, \, d, \, f,$ and $g$ symmetries.

\begin{sidewaystable}[!ht]
\caption{Convergence trends of magnetic fields $A_{\text{el}}$ (in MHz) and electric field gradients $q$ (in a.u.) of states in the [Xe]$5f^{14}5d^{10}6s^26p^3$ ground configuration of Bi with increase of correlation orbitals. AOs is referred to as the active orbitals that can be replaced with COs. COs stands for the set of correlation orbitals.}\label{HFS_convergence}
\begin{tabular}{ccccccccccccc}
\toprule%
                                                                      &&          \multicolumn{5}{c}{$A_{\text{el}}$}                                                                                                       &&          \multicolumn{4}{c}{$q$}                              \\
\cmidrule{3-7}\cmidrule{9-12}  AOs   &    COs    &   $^4\!S^o_{3/2}$   &   $^2\!D^o_{3/2}$   &   $^2\!D^o_{5/2}$    &   $^2\!P^o_{1/2}$   &   $^2\!P^o_{3/2}$   &&   $^4\!S^o_{3/2}$   &   $^2\!D^o_{3/2}$    &   $^2\!D^o_{5/2}$   &   $^2\!P^o_{3/2}$   \\
\midrule
\multicolumn{2}{c}{Two-configuration DHF}                                                                        &                                        87   &  -564  & 2917  &  11261 & 880 &&  1.446  &  7.628  &  -0.025  &  -9.003    \\
\makecell[c]{$6s^26p^3$ \\ $+~6s^26p6d^2$}                                                                     &   \textit{spdfg}             & -807  &  -1267 & 2349  &  11310 & 317 &&  2.048  &  5.519  &  -0.389  &  -8.157    \\
\makecell[c]{$5s^25p^65d^{10}6s^26p^3$ \\ $+~6s^26p6d^2$}                                         &  \textit{2spdfg}            & -717  &  -1397 & 2432  &  11665 & 293 &&  2.959  &  5.780  &  -0.110  &  -9.244    \\
\makecell[c]{$4s^24p^64d^{10}4f^{14}5s^25p^65d^{10}6s^26p^3$ \\ $+~6s^26p6d^2$}   &  \makecell[c]{\textit{3spdfg1hi} \\ \textit{2spdfg}}        & -577  &  -1470 & 2559  &  12046 & 342 &&  3.404  &  5.912  &  -0.109  &  -9.788    \\
\makecell[c]{$4s^24p^64d^{10}4f^{14}5s^25p^65d^{10}6s^26p^3$ \\ $+~6s^26p6d^2$}   &  \makecell[c]{\textit{4spdfg2hi} \\ \textit{2spdfg}}        & -580  &  -1513 & 2594  &  12230 & 341 &&  3.544  &  5.958  &  -0.106  &  -9.973    \\
\makecell[c]{$4s^24p^64d^{10}4f^{14}5s^25p^65d^{10}6s^26p^3$ \\ $+~6s^26p6d^2$}   &  \makecell[c]{\textit{5spdfg3h2i} \\ \textit{2spdfg}}      & -575  &  -1528 & 2613  &  12311 & 345 &&  3.614  &  5.986  &  -0.102  &  -10.072   \\
\makecell[c]{$4s^24p^64d^{10}4f^{14}5s^25p^65d^{10}6s^26p^3$ \\ $+~6s^26p6d^2$}   &  \makecell[c]{\textit{6spdfg4h2i} \\ \textit{2spdfg}}      & -570  &  -1536 & 2624  &  12358 & 347 &&  3.635  &  6.001  &  -0.101  &  -10.106   \\
\makecell[c]{$4s^24p^64d^{10}4f^{14}5s^25p^65d^{10}6s^26p^3$ \\ $+~6s^26p6d^2$}   &  \makecell[c]{\textit{7spdf6g4h2i} \\ \textit{2spdfg}}    & -583  &  -1543 & 2624  &  12380 & 343 &&  3.645  &  6.011  &  -0.101  &  -10.126   \\
\makecell[c]{$4s^24p^64d^{10}4f^{14}5s^25p^65d^{10}6s^26p^3$ \\ $+~6s^26p6d^2$}   &  \makecell[c]{\textit{8spd7f6g4h2i} \\ \textit{2spdfg}}  & -588  &  -1544 & 2624  &  12388 & 342 &&  3.651  &  6.018  &  -0.101  &  -10.139   \\
\makecell[c]{$4s^24p^64d^{10}4f^{14}5s^25p^65d^{10}6s^26p^3$ \\ $+~6s^26p6d^2$}   &  \makecell[c]{\textit{9spd7f6g4h2i} \\ \textit{2spdfg}}  & -588  &  -1544 & 2626  &  12394 & 342 &&  3.653  &  6.022  &  -0.101  &  -10.145   \\
\hline
\hline
\end{tabular}
\end{sidewaystable}

It should be emphasized that the VV CSFs generated from the $6s^26p6d^2$ reference configuration, such as $6s^26d^2nl$ and $6p6d^2nl n'l'$ are actually equivalent to those which would be generated from triple (T) and quadruple (Q) substitutions from $6s^26p^3$ to correlation orbitals $nl$ and $n'l'$, but with the specific occupations on the $6d$ subshell. However, including such TQ substitutions leads to large configuration spaces and often exceeds available computational (space and time) resources.
The numbers of CSFs are listed in Table~\ref{NCSF} for levels with given total angular momenta. To speed up the SCF procedures, we adopted the so-called perturbation-theory-based configuration interaction approach~\cite{Gustafsson2017,Dzuba2017c}, also called Zero-First approach~\cite{Jonsson2022,Jonsson2023} in the context of GRASP. In this approach, the off-diagonal matrix elements between SD-substitution CSFs are set to zero, and the Hamiltonian matrix has the form 

\begin{equation}
H = 
\left(
\begin{array}{c|cccccccc}
\\[-0.4cm]
\mathbf{H^{00}}                                &               &                    &   \mathbf{H^{off}}    &   &               \\[0.1cm]
\hline
\multirow{3}{*}{$\mathbf{H^{off}}$}  &  \ddots  &                     &      \mathbf{0}         &   &                \\
                                                          &               &                    &      \ddots                &   &                \\
                                                          &               &  \mathbf{0} &                                 &   &   \ddots  \\
\end{array}
\right)  \, . \nonumber 
\end{equation}
Here, $\mathbf{H^{00}}$ represents the subspace of the Hamiltonian matrix based on CSFs in the MR configurations, 
$\mathbf{H^{off}}$ the off-diagonal matrix elements between MR and VV+CV CSFs, 
and the dots --- the diagonal matrix elements between the VV+CV CSFs. 

\begin{table}[!ht]
\caption{The Number of CSFs (NCSF) for atomic state functions with $J=1/2, 3/2$, and $5/2$ in the [Xe]$5f^{14}5d^{10}6s^26p^3$ ground configuration of Bi.}
\label{NCSF}
\begin{tabular}{cccc}
\toprule
Models          &        $J=1/2$      &           $J=3/2$           &              $J=5/2$           \\
\midrule
SCF               &        \phantom{1}279704       &         1366881            &            1110122            \\
MR-CI            &       1105841      &         2019240            &            2603629            \\
CV3               &       1518248       &        2768982            &            3563664            \\
\bottomrule
\end{tabular}
\end{table}

As can be seen from Table~\ref{HFS_convergence}, nine layers of correlation orbitals (except $4h$ and $2i$, due to their tiny influence) were included in the final orbital set, in order to achieve convergence of the atomic hyperfine properties. The contribution from the TQ-substitution CSFs, i.e.~those generated from the $6s^2 6p6d^2$ configuration, 
saturated very fast, therefore only two layers of correlation orbitals were included for TQ substitutions. In order to distinguish these two substitution orders (SD vs TQ),
we split the lines 4-10 in the correlation orbitals column into two parts, indicating the different sets of correlation orbitals corresponding to two reference configurations, respectively. The calculated MFEs~($A_{\text{el}}$) and EFGs~($q$) of levels in the [Xe]$5f^{14}5d^{10}6s^26p^3$ configuration are presented in Table~\ref{HFS_convergence} for $^{209}$Bi~I. 

After generating the orbital set, we carried out large-scale RCI computations in order to take account of electron correlations systematically and to include the effects of the Breit interaction. At this stage, in the first step we included those off-diagonal matrix elements which were excluded in the SCF calculations. This step was labeled as MR-CI in Table~\ref{HFS_CI}. Furthermore, CV electron correlations with the $n=3$ core shell and the Breit interaction were taken into account. The corresponding contributions are presented in the rows CV$_3$ and ``Breit" of Table~\ref{HFS_CI}, respectively. Note that the configuration space in CV$_3$ is a merger of MR-CI and the CSFs generated by SrD substitutions from the $n=3$ core and from valence orbitals, i.e.~from $3s^23p^63d^{10}4s^24p^64d^{10}4f^{14}5s^26p^65d^{10}6s^26p^3$, with the maximal set of correlation orbitals. The effect of the CV$_3$ electron correlation is of the order of several percent on A$_{\text{el}}$ and $q$. In particular, it reaches 5.8\% for 
A$_{\text{el}}$ of the ground state.

\begin{sidewaystable}[!ht]
\caption{Magnetic fields $A_{\text{el}}$ (in MHz) and electric field gradients $q$ (in a.u.) of states in the [Xe]$5f^{14}5d^{10}6s^26p^3$ ground configuration of Bi. MR-CI: multi-reference configuration interaction computations based on the orbital basis presented above; CV$_{3}$: cove-valence correlation effect between electrons in the $n=3$ core shell and those in the valence; Breit: the corrections from the Breit interaction.}\label{HFS_CI}
\begin{tabular}{rrrrrrrrrrr}
\toprule
              &          \multicolumn{5}{c}{$A_{\text{el}}$}            &&          \multicolumn{4}{c}{$q$}                               \\
\cmidrule{2-6}\cmidrule{8-11}  Models     &   $^4\!S^o_{3/2}$   &   $^2\!D^o_{3/2}$   &   $^2\!D^o_{5/2}$   &   $^2\!P^o_{1/2}$   &   $^2\!P^o_{3/2}$   &&   $^4\!S^o_{3/2}$   &   $^2\!D^o_{3/2}$   &   $^2\!D^o_{5/2}$   &   $^2\!P^o_{3/2}$   \\
\midrule
MR-CI     &  $-$572     &  $-$1357  &  2691   &  13245    &  564   &&  3.081        &  7.199  &  $-$0.256  &  $-$10.689       \\
CV$_3$  &    32          &   $-$13     &    56     &    182       &  21    &&  0.060        &  0.110  &  $\sim$0    &  $-$0.178         \\
Breit        & $-$13       &  $-$32      &  $-$11  &   $-$63     &   2     &&  $-$0.069   &  0.050  &  0.003       &  0.023               \\
Final       & $-$554     &  $-$1338   &   3007  &  13365     & 587   &&  3.072        &  7.360  &  $-$0.253  &  $-$10.844        \\
\bottomrule
\end{tabular}
\end{sidewaystable}

To evaluate the accuracy of our calculated A$_{\text{el}}$ and $q$, the remaining electron correlation effects were estimated based on the MR-CI model. 
Firstly, we expanded the MR set by the $6s6p^36d$ configuration, since it has mixing coefficients greater than 0.05. The SD substitutions from this reference configuration are equivalent to the restricted triple substitutions from $6s^26p^3$. In order to control the computation scale, only one layer of correlation orbitals was used to generate these CSFs. This contribution is less than $-6$~MHz for A$_{\text{el}}$ and 0.01~a.u. for $q$. Furthermore, the influence of full triple substitutions from the $n=6$ valence shell was of the order of $-15$~MHz for A$_{\text{el}}$, and of the order of $-0.004$~a.u. for $q$, for the states with $J=3/2$. In comparison with results obtained with the MR-CI model, we speculated that triple and quadruple substitutions make opposite contributions to A. 
It has been observed in early papers by Engels~et~al~\cite{Eng:93a,Engetal:96a}, that the indirect effects of the double substitutions are opposite to those of single substitutions, and similarly with triple and quadrupole substitutions, respectively. 
Similar behavior has also been observed in our earlier calculations of hyperfine structures~\mrg{\cite{Godetal:97a, Bieron2009a, Bieetal:2018a}}. For $q$ the remaining electron correlation in the valence shell was negligible.

Secondly, we assessed the effect of electron correlation related to the $n=5$ core shell for the $J=3/2$ states. Double and triple substitutions from this shell were found to change A$_{\text{el}}$ in opposite directions, and the maximum net contribution reached 5\% for the ground state. On the other hand, the effect on $q$ was only a few thousandths.

Thirdly, it was inferred from the CV$_3$ model that the correlations between the electrons in the deeper n=1 and n=2 core shells and the valence shells might contribute a few percent to A$_{\text{el}}$ and $q$. 


To sum up, the computational uncertainties were $1\!\sim\!2\%$ for EFGs and of the order of 10\% for $A_{\text{el}}$.
Note that we did not consider the hyperfine anomaly in the present work. In addition, the calculated energy separations in the ground configuration differ from those in the NIST database~\cite{Kramida2025} by about 5\%. It revealed to some extent that the aforementioned estimate of the computational uncertainty was perhaps rather optimistic. Therefore, we finally assigned an uncertainty in the computation 5\% to the EFGs.

Using the final EFG result of the ground state listed in Table~\ref{HFS_CI}, $q=3.072$~a.u., we extracted the nuclear quadrupole moment of the $^{209}$Bi isotope from the latest measurement~\cite{Barzakh2021}. It gave a value of $\mathrm{Q}(^{209}\text{Bi}) = -422(22)$~mb. Our quadrupole moment is consistent with those recently obtained by Skripnikov \textit{et al.} with the Fock space coupled cluster method, $\mathrm{Q}(^{209}\text{Bi}) = -418(6)$~mb~\cite{Skripnikov2021}, by Wilman \textit{et al.} with semi-empirical atomic structure calculations, $\mathrm{Q}(^{209}\text{Bi}) = -446(15)$~mb~\cite{Wilman2021}, and by Dognon and Pyykk{\"o} from molecular structure calculations, $\mathrm{Q}(^{209}\text{Bi}) = -422(3)$~mb~\cite{Dognon2023}.

\section{Conclusions}
Combining our calculated EFG using the MCDHF and RCI methods implemented in the Grasp2018 package with the latest measurement on the electric quadruple hyperfine interaction constant, we obtained a new nuclear quadrupole moment of the $^{209}$Bi isotope, $\mathrm{Q}(^{209}\text{Bi})= -422(22)$~mb. The computational uncertainty presented in parentheses was estimated based on systematic analysis of the electron correlation effects on the EFGs of levels in the ground configuration for the neutral bismuth atom. This quadrupole moment, as well as others recently obtained by atomic and molecular methods, made a ``world average" value of $\mathrm{Q}(^{209}\text{Bi}) = -420(17)$~mb.

In addition, the computational model designed in this work was also used to revisit the nuclear octupole moment for $^{209}$Bi~\cite{Li2022e}. The resulting octupole moment agreed with that obtained from the nuclear structure calculation with inclusion of the nuclear core polarization and the correction to the nuclear electromagnetic current due to velocity-dependent interactions~\cite{Senkov2002}. It further confirmed the significance of many-body interactions among nucleons in the $^{209}$Bi isotope.

\bmhead{Acknowledgements}
This work was supported by the National Natural Science Foundation of China~(Grant Nos. 12474250, 11804090).

\bibliography{Bi_JG}

\end{document}